\documentclass[12pt]{JHEP3}
\input epsf

\newcommand{\bsigma}{{\mbox{\boldmath $\sigma$}}}

\newcommand{\be}{\begin{equation}}
\newcommand{\ee}{\end{equation}}
\newcommand{\bea}{\begin{eqnarray}}
\newcommand{\eea}{\end{eqnarray}}
\newcommand{\ba}{\begin{array}}

\newcommand{\ea}{\end{array}}

\def\bbox{{\,
\lower0.9pt\vbox{\hrule \hbox{\vrule height 0.2 cm
\hskip 0.2 cm \vrule height 0.2 cm}\hrule}\,}}
\newcommand{\dsl}{\pa \kern-0.5em /}

\newcommand{\nn}{\nonumber \\}

\def\Tr{{\rm Tr\,}}

\def\ds{\raise.15ex\hbox{/}\kern-.57em\partial}
\def\Ds{\,\raise.15ex\hbox{/}\mkern-13.5mu D}
\def\Tr{{\rm Tr}\,}
\def\n{\nonumber}
\newcommand{\beq}{\begin{equation}}
\newcommand{\eeq}{\end{equation}} \newcommand{\beqn}{\begin{eqnarray}}
\newcommand{\eeqn}{\end{eqnarray}}

\preprint{
AEI-2002-058\\
KIAS-P02045\\
hep-th/0207264\\
}

\title{\Large\bf Deformed Matrix Theories with ${\cal N}=8$ 
and Fivebranes in the PP Wave Background} 

\author{Nakwoo Kim\\
Albert-Einstein-Institut \\
Max-Planck-Institut f\"{u}r Gravitationsphysik \\
Am M\"{u}hlenberg 1, D-14476 Golm, Germany\\
kim@aei-potsdam.mpg.de}

\author{Kimyeong M. Lee  \\
School of Physics, Korea Institute for Advanced Study\\
207-43, Cheongryangri-Dong, Dongdaemun-Gu, Seoul 130-012, Korea\\
klee@kias.re.kr }

\author{Piljin Yi\\
School of Physics, Korea Institute for Advanced Study\\
207-43, Cheongryangri-Dong, Dongdaemun-Gu, Seoul 130-012, Korea\\
piljin@kias.re.kr}

\abstract{
M(atrix) theory is known to be mass-deformed in the pp-wave background
and still retains all 16 dynamical supersymmetries. We consider generalization 
of such deformations on super Yang-Mills quantum mechanics (SYQM) 
with less supersymmetry. In particular this
includes ${\cal N}=8$ $U(N)$ SYQM with a single adjoint and any number 
of fundamental hypermultiplets, which is a pp-wave deformation
of DLCQ matrix theory of fivebranes. With $k\ge 1$ fivebranes,
we show that a rich vacuum structure exists, with many continuous 
family of solutions that preserve all dynamical supersymmetries.
The vacuum moduli space contains copies of $CP^{k-1}$ of 
various sizes.
}

\begin{document}

\section{Introduction}

It was discovered recently by Berenstein, Maldacena, and Nastase (BMN) 
\cite{bmn}
that ${\cal N}=16$ supersymmetric Yang-Mills quantum mechanics (SYQM) 
admits a massive deformation without sacrificing any of the
supersymmetries. Instead, the supersymmetry algebra is deformed in such
a way that Hamiltonian no longer commutes with supercharges, implying that
supersymmetry transformation itself is explicitly time dependent. 

Usual SYQM with 16 supersymmetries \cite{QM}
had played a crucial role in uncovering 
the 11-dimensional nature of M-theory \cite{M}. 
The only known proposal for quantum 
formulation of M-theory, known as M(atrix) theory \cite{bfss}, 
is simply a large $N$ 
limit of $U(N)$ SYQM with 16 supersymmetries, which can be thought of as
large $N$ dynamics of D0 branes \cite{bound} 
or alternatively as a regularized dynamics of
supermembranes \cite{dWHN}.
In this regard, the deformed SYQM of BMN may be thought of as 
reformulation of M theory in a particular curved background which is
a Penrose limit of either $AdS_7\times S^4 $ or $AdS_4 \times S^7$
\cite{ppwaves}. 
Again, deformed SYQM may be considered as dynamics of D0 particles 
or as  regularized dynamics of membranes in such a background \cite{DSJvR}. 
The spacetime in question is often referred to as M-theory pp-wave
background. Naturally one would like to ask what we can learn about M-theory 
by studying such mass-deformed version of M(atrix) theory. Some
initial steps toward understanding quantum aspect of this deformed theory
have been taken in Ref.~\cite{plefka,jhp}.

One of more 
puzzling objects in M(atrix) theory is the fivebrane. Original proposal 
of M(atrix) theory apparently does not contain fivebranes as a solution, 
rather one must put them in by hand. For longitudinal fivebranes, this is
achieved by introducing additional fields, namely fundamental hypermultiplets, 
in SYQM, thereby reducing dynamical supersymmetry by half to ${\cal N}=8$
\cite{Aharony}. 
This quantum SYQM with 8 supersymmetries can also be used to study
fivebranes themselves, and may be considered as an analog of M(atrix) 
theory for worldvolume dynamics of fivebranes, namely a 
Discrete-Lightcone-Quantization (DLCQ) description.
In this note, we consider general ${\cal N}=8$ SYQM and their
mass deformation analogous to BMN's, thereby
initiating the inquiry as to whether such deformed 
M(atrix) theory with fivebranes allows better understanding of
fivebranes in particular.

Reducing supersymmetry to ${\cal N}=8$ also involves reducing R-symmetries.
Prior to mass deformation, ${\cal N}=16$ SYQM are equipped with $SO(9)$
R-symmetry coming from spatial rotations. Reduction to ${\cal N}=8$ breaks
up this to $SU(2)\times SO(5)$, where $SO(5)$ is associated with directions
orthogonal to fivebranes. In the deformed SYQM of BMN with ${\cal N}=16$,
however, $SO(9)$ is already reduced to a smaller $SO(3)\times SO(6)$ 
R-symmetry by the four-form flux. Thus, depending on how we embed fivebranes,
we will find different theories, not all of which may be supersymmetric.

In terms of fivebrane in the pp-wave background, this is because  the 
background itself comes from $AdS_7\times S^4$ (or $AdS_4 \times S^7$),
a near-horizon limit of fivebranes, so not all orientations of the secondary 
fivebrane are equivalent. After some experiment with spinors, we discover 
that there is one particular subset of orientation that is amenable to 
reduction of supersymmetry and thus to introduction of additional 
supersymmetric fivebranes. Let us describe this choice in some detail.
We start with $AdS_7\times S^4$ such that $1+10$ directions are distributed
as follows,
\beqn
x^{0,1,2,3,4,8,9}&\rightarrow &AdS_7,\\
x^{5,6,7,10}&\rightarrow & S^4.
\eeqn
Perform the pp-wave limit of this background by taking $x^\pm=x^0\pm x^{10}$.
The BMN deformed M(atrix) theory then has $SO(3)\times SO(6)$ R-symmetry
where $SO(6)$ rotates $x^{1,2,3,4,8,9}$ while $SO(3)$ rotates $x^{5,6,7}$.
Reduction of dynamical supersymmetry to ${\cal N}=8$ is achieved in part
by breaking up $x^{1,2,3,\dots,9}$ into five coordinates associated
with vector multiplets and four coordinates associated with hypermultiplets.
In this note, we take the following decomposition,
\beqn
x^{1,2,3,4,5}&\rightarrow & \hbox{vector},\n\\
x^{6,7,8,9}&\rightarrow & \hbox{hyper}.
\eeqn
The resulting SYQM will have 
\beq
SO(4)_{1234}\times SO(2)_{67-89},
\eeq
as the R-symmetry. $SO(4)$ rotates directions 1,2,3,4, among themselves
while $SO(2)$ is a simultaneous and opposite rotation of planes 6,7 and 8,9.

This choice of decomposition seems special in the following sense. Recall 
that in the deformed SQYM, the supercharge does not commute with Hamiltonian
and the supermultiplet does not imply degeneracy. In particular, bosonic
and fermionic degrees of freedom come with different excitation masses.
Only after summing over all vacuum energies, one find cancellation among
them giving a supersymmetric ground state of zero energy. The above
decomposition into vector and hyper is unique up to irrelevant rotation, 
in that this cancellation of vacuum energy holds in each sector.
This can be seen easily once we recall
how the cancellation occurred in full ${\cal N}=16$ BMN case,
\beq
3\times \frac{|\mu|}{3}+6 \times \frac{|\mu|}{6} -8
\times \frac{|\mu|}{4}=0,
\eeq
where $3$ is from $x^{5,6,7}$, 6 is from $x^{1,2,3,4,8,9}$,
and 8 is from 16 real fermions. With the above decomposition into vector 
and hyper then, this is rewritten as
\beq
\left(1\times \frac{|\mu|}{3}+4\times \frac{|\mu|}{6}
-4\times \frac{|\mu|}{4}\right)+
\left(2\times \frac{|\mu|}{3}+2\times \frac{|\mu|}{6}
-4\times \frac{|\mu|}{4}\right)=0,
\eeq
where each group gives zero separately. The first group represents
vacuum energy from a vector multiplet, and the second represents
vacuum energy from a hypermultiplet. In particular, this already suggests that 
we may truncate to vector part only or add additional hypermultiplets 
in the SYQM without destroying supersymmetry.

In section 2, we will start with mass deformed ${\cal N}=16$ SYQM 
and rewrite it in ${\cal N}=8$ language. By either
adding or subtracting hypermultiplet into the latter setup, we show that
general ${\cal N}=8$ SYQM also admits massive deformation with all 8 
supercharges remaining. In section 3, we outline how ${\cal N}=16$ 
superalgebra is truncated to   ${\cal N}=8$ with emphasis on central charge.
In section 4, we concentrate on the case with
fundamental hypermultiplets which is DLCQ of fivebranes in pp-wave background,
and isolate classical ground states that preserves all 8 dynamical 
supersymmetries. These solutions are analogous to the fuzzy sphere 
solutions of BMN which are essentially giant gravitons. Unlike the BMN
case, however, we find that the solutions come in a continuous family
typically parametrized by a $CP^n$ vacuum moduli space. In section
4 we conclude with remarks on future direction of research.

\section{Mass Deformed ${\cal N}=8$ SYQM}

\subsection{10 dim Yang-Mills and Deformed Matrix Theory in Real
Representation} 

Here we review the mass deformed SYQM of BMN and set up notations
and conventions. With signature $(-,++,...,+)$ and
$\bar{\lambda}^a=(\lambda^a)^T\Gamma^0$, we start with supersymmetric
Yang-Mills theory in ten dimensions,
\beq
L =\Tr \left\{ -\frac{1}{4} F^2 - \frac{i}{2} \bar{\Psi}\Gamma^P D_P
\lambda \right\},
\eeq
where $A_P$ is hermitian and $ D_P \Psi = \partial_P\lambda
-i[A_P, \Psi]$.

The 10 dimensional Gamma matrices can be chosen
to be real 
\beqn
&& \Gamma^0=1 \otimes i\sigma_2 ,\n\\
&& \Gamma_I =\bar \gamma_I\otimes \sigma_1 \;\; (I=1,2...9),\n\\
&& \Gamma^{11} = 1\otimes \sigma_3,
\eeqn
where $\bar \gamma_I$ is the symmetric real $16\times 16$ gamma matrices
of $SO(9)$. The gaugino field is Majorana and is also chiral,
\beq 
\Gamma_{11}
\Psi=\Psi.
\eeq
Note that $\Gamma^0\Gamma_I \Psi=\bar \gamma_I \Psi$. Because of this,
we may as well regard $\bar\gamma_I$ as $32\times 32$ matrices
\beq
\bar\gamma_I=\Gamma^0\Gamma_I
\eeq
with the understanding that the spinor $\Psi$ is restricted to be
Majorana and chiral. This latter definition of $\bar\gamma_I$
naturally extends to complex representation of spinors we 
will later adopt.

Supersymmetric transformation is
\beqn
&& \delta A_I = -i \bar{\epsilon} \Gamma_I \Psi , \n\\
&& \delta \Psi = \frac{1}{2} F_{IJ}
\Gamma^{IJ}\epsilon.
\eeqn
Let us dimensionally reduce this field theory to a quantum mechanics by
allowing time-dependence only. With new notations, $X_I= A_I$ and $A_0$, 
the above Lagrangian becomes \cite{QM}
\beq
L=  \Tr \left\{ \sum_I \frac{1}{2}(D_0X_I)^2 +
\frac{1}{4}\sum_{IJ}[X_I,X_J]^2 + \frac{i}{2}\Psi^T D_0\Psi
-\frac{1}{2} \Psi^T \bar \gamma_I[X_I, \Psi ] \right\}.
\eeq
The susy  transformation of the quantum mechanics becomes
\beqn
&& \delta A_0 = i\Psi^T \epsilon,  ,\n\\
&& \delta X_I= i\Psi^T\bar \gamma_I \epsilon,   \n \\
&& \delta \Psi = D_0 X_I \bar \gamma_I \epsilon -\frac{i}{2}
[X_I,X_J]\bar \gamma_{IJ} \epsilon .
\eeqn
BMN \cite{bmn} deformed this quantum mechanics by adding the following set of
terms, with one mass parameter $\mu$,
\beqn
\Delta L&=&\frac{1}{2}\Tr\left( -\left(\frac{\mu}{3}\right)^2
\sum_{a=5,6,7} (X_a)^2 -\left(\frac{\mu}{6}\right)^2 
\sum_{s=1,2,3,4,8.9}(X_s)^2 \right) \n \\
&+&\frac{1}{2}\Tr\left( \frac{2i\mu}{3} 
\epsilon_{abc} X_a X_b X_c -\frac{i\mu}{4} \Psi^T\bar \gamma_{567}\Psi
  \right),
\eeqn
upon which the susy transformation get deformed to
\beqn
\delta A_0 &=& i\Psi^T \epsilon ,   \n \\
\delta X_I&=& i\Psi^T \bar \gamma_I \epsilon , \n \\
\delta \Psi &=& \left( D_0 X_I \bar \gamma_I  -\frac{i}{2}
[X_I,X_J]\bar \gamma_{IJ} \right), \n\\
&+&\left(\frac{\mu}{3}\sum_{a=5,6,7} X_a\bar \gamma_a
\bar \gamma_{567} -\frac{\mu}{6}\sum_{s=1,2,3,4,8,9} X_s\bar \gamma_s 
\bar \gamma_{567}
\right) \epsilon.
\eeqn
The deformed Lagrangian is invariant under this supersymmetry if
and only if we force the following explicit time-dependence of the
transformation parameter $\epsilon$,
\beq
\epsilon(t) = e^{-\frac{\mu}{12}\bar\gamma_{567}t} \epsilon(0) , \label{time}
\eeq
which makes the superalgebra quite unconventional. In particular,
the supercharges does not commute with Hamiltonian, but raises
or lower energy by $\mu/12$ unit. The notion of ``supermultiplet''
no longer implies degeneracy in this deformed superalgebra.
Note that we chose $SO(3)$ part of $SO(9)$ to lie along direction $x^{5,6,7}$
for later convenience.

\subsection{Complex Representation and Undeformed ${\cal N}=8$ SYQM}

In order to reduce to SYQM with ${\cal N}=8$ supersymmetry and introduce
other kind of supermultiplet, we break up the $SO(9)$ spinor
in terms of tensor product of $SO(4)$ and $SO(5)$ spinors. 
First introduce $e_\mu= (\bsigma,i)$ and $\bar{e}_\mu =
(\bsigma,-i)$ and define the Euclidean SO(5) gamma 
matrices\footnote{to be distinguished from 
$\{\bar \gamma_I\}=\{\bar \gamma_s, \bar\gamma_a\}$}
\beqn
&& \gamma_\mu = \left(\begin{array}{cc} 0 & e_\mu \\ \bar{e}_\mu & 0 
                   \end{array}\right) , \n \\
&& \gamma_5 = \left(\begin{array}{cc} 1 & 0 \\ 0 & -1 
                   \end{array}\right) ,\label{5d}
\eeqn
which implies
\beqn
&&\gamma_1 = \sigma_1\otimes \sigma_1  , \n\\
&&\gamma_2 = \sigma_2\otimes \sigma_1  , \n\\
&&\gamma_3=\sigma_3 \otimes \sigma_1  , \n\\
&&\gamma_4 = -1\otimes \sigma_2 , \n \\
&&\gamma_5 = 1\otimes \sigma_3.
\eeqn
Note that $\gamma_5=\gamma_1\gamma_2\gamma_3\gamma_4$. With this, 
ten dimensional Dirac matrices can be written in a complex form
as
\beqn
&& \Gamma^0= 1\otimes 1\otimes i\sigma_2 \otimes 1\otimes 1  , \n\\
&& \Gamma_\mu = \gamma_\mu \otimes \sigma_1\otimes 1\otimes 1 ,
\;\; ( \mu=1,2...,5)  , \n\\
&& \Gamma_{5+\mu} = 1\otimes 1\otimes \sigma_3\otimes \gamma_\mu ,
\;\; (\mu=1,2,3,4)  , \n\\
&& \Gamma_{11} = 1\otimes 1\otimes \sigma_3\otimes 1\otimes \sigma_3,
\label{gamma11} 
\eeqn
with $1$ denoting the $2\times 2$ identity matrix.
As before, $\Gamma_{11}= \Gamma^0\Gamma_1...\Gamma_{10}$. 

Since we
abandoned real $SO(9)$ matrices, we must impose  Majorana condition.
Defining complex conjugation by
\beq
\Psi_c= B\Psi^* ,
\eeq
with the matrix $B$ such that
\beq
B\Gamma^I B^{-1} = (\Gamma^{I})^* .
\eeq
The Majorana condition is then $\Psi_c=\Psi$.  ($C= B\Gamma^0$ and
$C\Gamma^I C^{-1} = -(\Gamma^I)^T$ with $\Psi_c = C\bar{\Psi}^T$.)  In
our matrix convention
\beq
B = -\sigma_2\otimes\sigma_3\otimes 1\otimes\sigma_2\otimes \sigma_3 .
\label{B}
\eeq
The above Yang-Mills Lagrangian and supersymmetry work fine if we
insist $\lambda$ and $\epsilon$ are both chiral and Majorana.
In this notation, one starts with a 32-component complex spinor,
then the Majorana condition together with $\Gamma_{11}\Psi=\Psi$
reduces it to 16 real in effect.

We will decompose the spinor $\Psi$ into two eight-component complex
spinors, 
\beq
\lambda^i_\alpha,\qquad \chi^i_\alpha
\eeq
which are 
respectively chiral and anti-chiral under $\Gamma^0\Gamma^1\cdots 
\Gamma^5$.  Similarly, supersymmetry parameter $\epsilon$ is split
into $\epsilon_i$ and $\zeta_i$. The 10-dimensional chirality 
condition then implies that $\lambda_i$ and $\epsilon_i$
are of $(+,+)$ eigenvalues under the two $\sigma_3$ factors in
the above form of $\Gamma_{11}$
while $\chi_i$ and $\zeta_i$ are of $(-,-)$. 
Eventually we will keep the supersymmetries generated by $\epsilon_i$'s 
only. The Majorana condition becomes symplectic Majorana conditions
\beq
\label{syma}
\lambda_i = -\tilde{B} (\sigma_2)_{ij}\lambda^*_j ,\qquad
\chi_i = +\tilde{B} (\sigma_2)_{ij}\chi^*_j ,
\eeq
where $\tilde{B} = \sigma_2\otimes \sigma_3$ satisfies $\tilde{B}
\gamma^\mu \tilde{B}^{-1} = \gamma^{\mu*}$ with $\mu=1,2,...5$. Note
that the indices $i,j=1,2$ belong to the fourth factor of spinor
decomposition (\ref{gamma11}); $\sigma_2$ in the preceding equation
is the fourth entry in the definition of $B$ in eq.~(\ref{B}) above.

For fermions associated with hypermultiplet, the gauge representation
could be either real or complex, so we may as well take one of
$\chi_i$'s and discard the other. We will take $\chi=\chi_1$ as the
fermion of the hypermultiplet. Similarly we introduce the two complex 
scalar belonging to the hypermultiplet.
\beq
\sum_{\mu=1}^4 X_{\mu+5}e_\mu = \left( \begin{array}{cc}
                                   \bar{y}_1 & y_2  \\
				   \bar{y}_2 & -y_1 
				\end{array} \right),
\eeq
where $y_1= X_8-iX_9$ and $y_2=X_6-iX_7$.

With the above decomposition, the pure ${\cal N}=8$ SYQM without mass 
deformation has
\beq
L_0=\Tr\left( \frac{1}{2} \sum_{\mu=1}^5 ( D_0 X_\mu)^2 + 
\frac{1}{4} \sum_{\mu , \nu}^5 [X_\mu,X_\nu]^2 + \frac{i}{2}
\lambda_i^\dagger D_0 \lambda_i  -\frac{1}{2} \lambda_i^\dagger
\gamma_\mu [X_\mu,\lambda_i] + \frac{1}{2} {\bf D}^2 
\right) ,\label{pure}
\eeq
which is invariant under
\beqn
&& \delta A_0 = i\lambda_i^\dagger \epsilon_i  , \n\\
&& \delta X_\mu=i\lambda_i^\dagger \gamma_\mu \epsilon_i  , \n\\
&& \delta \lambda_i = D_0 X_\mu \gamma_\mu \epsilon_i - \frac{i}{2}
[X_\mu, X_\nu]\gamma_{\mu \nu} \epsilon_i + i {\bf D}\cdot
\bsigma_{ij}\epsilon_{j}  , \n\\
&& \delta {\bf D} =  - \epsilon^\dagger_i \bsigma_{ij} D_0\lambda_j -
i\epsilon^\dagger_i \bsigma_{ij} \gamma_\mu[X_\mu,\lambda_j] . \label{susy0}
\eeqn
Adding an adjoint hyper matter field is a matter of rewriting ${\cal N}=16$ 
undeformed SYQM in terms of the above decomposition
\beqn
{ L}_{adj} &=&  {\hskip -3mm}  \Tr \left( \frac{1}{2} D_0
\bar{y}_i D_0 y_i  + \frac{1}{2} [X_\mu,\bar{y}_i]
[X_\mu,y_i]   + 
\frac{1}{2} {\bf D} \cdot \bsigma_{ij} [\bar{y}_j,y_i]
\right.    \nonumber \\ 
&&  \quad\left.  + 
i\chi^\dagger D_0\chi+ \chi^\dagger \gamma_\mu [X_\mu,\chi] +
\lambda_i^\dagger [\bar{y}_i,\chi]+\chi^\dagger[y_i,\lambda_i]
 \right) ,\label{adj}
\eeqn
The combined action is invariant under
the above susy transformation if we also transform,
\beqn
&& \delta\bar{y}_i= -2i\chi^\dagger \epsilon_i , \n \\
&& \delta \chi=-D_0 y_i \epsilon_i -i\gamma_\mu [X_\mu, y_i]\epsilon_i 
\label{susy1}
\eeqn
Adding the fundamental matter field $q,\psi$ in the $\bar{N}$ dimensional 
representation becomes
\beqn
{ L}_{fund} &=& \Tr \left( \frac{1}{2} D_0
\bar{q}_i D_0 q_i - \frac{1}{2}
X_\mu	\bar{q}_iq_i X_\mu    + 
\frac{1}{2} {\bf D} \cdot \bsigma_{ij} \bar{q}_j q_i
\right.    \n\\ 
&&  \quad \left.  + 
i\psi^\dagger D_0\psi -  \psi^\dagger \gamma_\mu \psi X_\mu +
\lambda_i^\dagger \bar{q}_i\psi+\psi^\dagger q_i\lambda_i \right) .\label{fund}
\eeqn
The susy transformation extends to
\beqn
&& \delta \bar{q}_i = -2i \psi^\dagger \epsilon_i , \n \\
&& \delta \psi = -D_0q_i \epsilon_i +i q_i X_\mu\gamma_\mu  \epsilon_i .
\label{susy2}
\eeqn

\subsection{Deformation of Pure ${\cal N}=8$ SYQM}

%
%

The above decomposition of spinors gives the relations
\beq
\bar\gamma^{567} = \gamma^5\otimes \sigma_3\otimes i\sigma_3\otimes 1.
\eeq
When acting on $\epsilon_i$'s and on $\lambda_i$'s, the $\sigma_3$ in
the middle takes $+1$ eigenvalue, while $i\sigma_3$ acts on $i=1,2$
indices. Thus we find that
\beq
\label{time24}
\epsilon_i(t) = 
\left(e^{-\frac{i\mu}{12} \gamma_5 \otimes \sigma_3}\right)_{ij}
(\epsilon_0)_j.
\eeq
Let us introduce one more notation. Since $\lambda_{1,2}$ are related
by reality condition and so are $\epsilon_{1,2}$, define
\beqn
&&\lambda_1= \lambda , \n \\
 &&\epsilon_1 = -i\epsilon,
\eeqn
from which it follows
\beqn
&& \lambda_2 = -i\tilde{B} \lambda^*  , \n\\
&& \epsilon_2 = \tilde{B} \epsilon^*,
\eeqn
and is consistent with the above time-dependence of $\epsilon_i$'s.

The vector multiplet Lagrangian becomes 
\beq
{L}_0 = \Tr \left( \sum_{\mu=1}^5\frac{1}{2} (D_0 X_\mu)^2 +  
\sum_{\mu, \nu}^5 \frac{1}{4} [X_\mu,X_\nu]^2 +
i\lambda^\dagger D_0 \lambda -  \lambda^\dagger
\gamma_\mu [X_\mu,\lambda] +\frac12 {\bf D}^2\right). \label{pure'}
\eeq
The additional piece of Lagrangian that deforms a pure ${\cal N}=8$ SYQM is
then
\beq
\Delta L_0 = \Tr \left( -\frac{1}{2} \left(\frac{\mu}{6}\right)^2
\sum_{\mu=1}^4 X_\mu^2 -\frac{1}{2} \left(\frac{\mu}{3}\right)^2
X_5^2  + \frac{\mu}{4} \lambda^\dagger \gamma_5 \lambda \right) .
\label{pure-1}
\eeq
The Lagrangian $L_0 +\Delta L_0$ is invariant under the susy 
\beqn
 \delta  A_0&=&(\lambda^\dagger \epsilon +\epsilon^\dagger \lambda) , \n
\\
 \delta  X_\mu &=& (\lambda^\dagger \gamma_\mu  \epsilon
+\epsilon^\dagger \gamma_\mu \lambda)   , \n\\
\delta \lambda &=& -i  D_0X_\mu \gamma_\mu
\epsilon - \sum_{\mu , \nu}^5 \frac{1}{2} [X_\mu,X_\nu]
\gamma_{\mu  \nu}\epsilon  \n\\
&& -\frac{\mu}{6} \sum_{\mu=1}^4
X_\mu\gamma_\mu \gamma_5 \epsilon + \frac{\mu}{3} X_5 \epsilon,
\label{susy0-1}
\eeqn
with
\beq
\epsilon(t) = e^{-\frac{i\mu}{12} \gamma_5 t} \epsilon_0.
\eeq
This deformed pure ${\cal N}=8$ SYQM theory has 16 supersymmetry out of 
which eight is linearly realized, or dynamical, and eight is  nonlinearly
realized, or kinematical.

Note that the oscillator mass of the bosonic and fermionic degrees of freedom
are again different from each other. Of five scalars, one is of mass $|\mu/3|$,
namely $X^5$, while the remaining four are of mass $|\mu/6|$. As before, all 
four complex fermions are of mass $|\mu/4|$. The vacuum energy is then,
\beq
\hbox{dim}(U(N))\times \left(\frac{|\mu|}{3}+4\times \frac{|\mu|}{6}
-4\times \frac{|\mu|}{4}\right)=0 ,
\eeq
as promised. It has been shown recently that such vacuum energy is
protected from correction by supersymmetry in ${\cal N}=16$ 
setting \cite{jhp}, which should also hold in ${\cal N}=8$ case.

\subsection{BMN in Complex Notation:
Deformation of ${\cal N}=8$ SYQM with an Adjoint Hypermultiplet}

The hypermultiplet part of the deformed Lagrangian can be written
\beq
\Delta {L}_{adj} = \Tr \left
( -\frac{1}{2} \left(\frac{\mu}{6}\right)^2 \bar{y}_1 y_1 
-\frac{1}{2}\left(\frac{\mu}{3}\right)^2 \bar{y}_2y_2  
-\frac12 \mu X_5[\bar{y}_2, y_2] -
\frac{\mu}{4} \chi^\dagger \gamma_5 \chi \right) .\label{adj-1}
\eeq
As usual, susy transformation of vector multiplet is changed as
follows
\beq
\Delta \delta \lambda = iD^3\epsilon_1 + (iD^1+D^2)\epsilon_2 ,
\eeq
when hypermultiplets are present. In addition the hypermultiplet fields
must transform as
\beqn
&& \delta \bar{y}_i = -2i \chi^\dagger \epsilon_i(t)   , \n\\
&& \delta \chi = -D_0y_i \epsilon_i - i \sum_{\mu=1}^5[X_\mu,
y_i]\gamma_\mu \epsilon_i + \frac{i\mu}{6}\gamma_5 y_1 \epsilon_1 +
\frac{i\mu}{3} \gamma_5 y_2 \epsilon_2 . \label{susy1-1}
\eeqn
Recall that $\epsilon_1= -i\epsilon $ and $\epsilon_2 = -i\tilde{B}
\epsilon_1^* $. 

This completes rewriting of BMN's deformed M(atrix) theory in terms
of ${\cal N}=8$ supersymmetry. While obvious, it is worthwhile to point
out that vacuum energy contribution from hypermultiplet degrees of
freedom cancel among themselves. Bosons split into two real with
mass $|\mu/3|$ and two real with $|\mu/6|$ while all four complex
fermions are of mass $|\mu/4|$, so that
\beq
\hbox{dim}(U(N))\times \left(2\times \frac{|\mu|}{3}+2\times \frac{|\mu|}{6}
-4\times \frac{|\mu|}{4}\right)=0.
\eeq

\subsection{Deformation with Fundamental Hypermultiplet}

Finally we are ready to add other hypermultiplets. To be specific,
let us add fundamental hypermultiplets for $U(N)$ while for other
multiplets, this should carry over  verbatim. Let $q_i$ complex scalars
and $\psi$ complex spinor in representation $\bar N$ as before,
\beq
\Delta {L}_{fund} = \Tr \left
( -\frac{1}{2}\left(\frac{\mu}{3}\right)^2 \bar{q}_2 q_2 
-\frac{1}{2} \left(\frac{\mu}{6}\right)^2 \bar{q}_1 q_1
-\frac12 \mu X_5\,\bar{q}_2 q_2 -
\frac{\mu}{4} \psi^\dagger \gamma_5 \psi \right) .
\label{fund-1}
\eeq
The susy transformation extends to
\beqn
&& \delta \bar{q}_i = -2i \psi^\dagger \epsilon_i(t)   , \n\\
&& \delta \psi = -D_0q_i \epsilon_i + i \sum_{\mu=1}^5
q_iX_\mu \gamma_\mu \epsilon_i + \frac{i\mu}{6}\gamma_5 q_1 \epsilon_1 +
\frac{i\mu}{3} \gamma_5 q_2 \epsilon_2 .\label{susy2-1}
\eeqn
As in the above adjoint hypermultiplet, the vacuum energy cancels
among each hypermultiplet
\beq
N \times \left(2\times \frac{|\mu|}{3}+2\times \frac{|\mu|}{6}
-4\times \frac{|\mu|}{4}\right)=0.
\eeq

\section{Deformed ${\cal N}=8 $ Superalgebra}

Let us start by rewriting deformed ${\cal N}=16$ superalgebra of BMN 
in a form suitable for the Dirac matrices of section 2.2. Since the Dirac 
matrix is now complex, fermions and supercharges are all necessarily 
complex. The reality property is achieved by imposing a 
Majorana condition $\Psi=B\Psi^*$. Thus all 16 supercharge must satisfy,
\beq
{\cal Q}=B({\cal Q}^\dagger)^T,
\eeq
where the transposition is for the spinor indices only. To be definite,
let us write their explicit form,\footnote{$\bar\gamma_I$'s in this 
section differ from those introduced in section 2.1. Instead, we have
$\bar\gamma_I=\Gamma^0\Gamma_I $ in terms of complex
Dirac matrices of (\ref{gamma11}). }
\beqn
{\cal Q}_M &=& \left(P_I\bar\gamma_I +\frac{i}{2}[X_I,X_J]\bar\gamma_{IJ}
-\frac{\mu}{3}X_a\bar\gamma_a\bar\gamma_{567}-\frac{\mu}{6} 
X_s\bar\gamma_s\bar\gamma_{567}\right)_{MK}\Psi_K ,\nn
{\cal Q}^\dagger_K &=& \Psi^\dagger_M \left(P_I\bar\gamma_I 
-\frac{i}{2}[X_I,X_J]\bar\gamma_{IJ}
+\frac{\mu}{3}X_a\bar\gamma_a\bar\gamma_{567}-\frac{\mu}{6} 
X_s\bar\gamma_s\bar\gamma_{567}\right)_{MK},
\eeqn
with $SO(9)$ spinor indices $M$ and $K$. As usual the canonical
commutators are $[P_I,X^J]=-i\delta_I^J$. 
The anticommutators of the supercharges are\footnote{Our convention
is such that \beq {\cal L}_{IJ}=X_IP_J-X_JP_I+\cdots. \eeq}
\beqn
\{{\cal Q}_M,{\cal Q}^\dagger_K\}&=&\delta_{MK}2 {\cal H}
-\frac{\mu}{3}\left(\bar\gamma_{ab}\bar \gamma_{567}\right)_{MK}{\cal L}_{ab} 
+\frac{\mu}{6}\left(\bar\gamma_{st}\bar\gamma_{567}\right)_{MK}{\cal L}_{st}\nn
&+& i(\bar\gamma^I)_{MK}
\Tr\left[X^I{\cal G}\right], \label{qbq}
\eeqn
where $\cal G$ is the Gauss constraint. As before, $a,b$ run over 5,6,7, while 
$s,t$ run over 1,2,3,4,8,9. Supercharges do not commute with Hamiltonian
in these mass deformed theories, but raise or lower energy as follows,
\beq
[{\cal H},{\cal Q}^\dagger_K]=-i\frac{\mu}{12}{\cal Q}^\dagger_M
\left(\bar\gamma_{567}\right)_{MK}.
\eeq
As a matter of convenience, we regard this theory as  ${\cal N}=8$ with 
a single adjoint hypermultiplet, and consider a subset of 8 supercharges.
Recall that 
\beq
\Gamma_{11}=1\otimes 1\otimes \sigma_3\otimes 1\otimes \sigma_3 \label{11},
\eeq
with respect to which all fermions and thus all 16 supercharges must
be of +1 eigenvalue. Thus 16 supercharges are split into two classes;
\beq
{\cal Q} = Q \oplus Q',
\eeq
where
$Q_{\alpha}^i$ are $(+,+)$ eigenvalues with respect to
two $\sigma_3$ factors of $\Gamma_{11}$
and $(Q')_\alpha^i$ are of $(-,-)$ eigenvalues. 
The reduction we took involves keeping the former.

Reduction to ${\cal N}=8$ 
is a matter of picking out $(+,+)$ eigensectors under
$\Gamma_{11}$. This means that of various combinations of Dirac matrices
appearing in the right hand side of Eq.~(\ref{qbq}), 
only those that are diagonal under the decomposition ${\cal Q} = 
Q \oplus Q'$ may survive the truncation. We list all such combinations,
and their truncated form as $(4\times 4)\otimes (2\times 2)$ ,
\beqn
\bar\gamma_{67}\bar\gamma_{567} &\rightarrow& \gamma_5\otimes (-1),\n\\
\bar\gamma_{89}\bar\gamma_{567} &\rightarrow& \gamma_5\otimes (+1)\n\\
\bar\gamma_{\mu\nu}\bar\gamma_{567} &\rightarrow& 
\gamma_{\mu\nu}\gamma_5\otimes i\sigma_3\qquad\; (\mu,\nu=1,2,3,4),\n\\
\bar\gamma_\mu &\rightarrow& \gamma_\mu\otimes \sigma_3
\qquad\qquad(\mu=1,2,3,4,5),
\eeqn
in order of their appearance in (\ref{qbq}). The reduced superalgebra is then,
\beqn
\{{Q}_{\alpha i},{Q}^\dagger_{\beta j}\}&=&\delta_{ij}
\delta_{\alpha\beta}\,2 {H} \nn
&+&\mu\,\delta_{ij} (\gamma^5)_{\alpha\beta} 
\, [\,T\,]
+\frac{\mu}{3}\,\delta_{ij} (\gamma^5)_{\alpha\beta}\, [\,S\,] 
+ \frac{\mu}{6}\sum_{\mu,\nu=1}^4
\,i(\sigma_3)_{ij} (\gamma^{\mu\nu}\gamma^5)_{\alpha\beta}\,
\left[\,L_{\mu\nu}\,\right] \nn
&+&\sum_{\mu=1}^5 i(\sigma_3)_{ij} (\gamma^\mu)_{\alpha\beta}\,
\Tr\left[X^\mu{\cal G}\right], \nn
\eeqn
and
\beq
[{H},{Q}^\dagger_{\beta j}]=\frac{\mu}{12}{Q}^\dagger_{\alpha i}
(\gamma^5)_{\alpha\beta}(\sigma_3)_{ij},
\eeq
with
\beqn 
S&\equiv &\frac12\left[L_{67}-L_{89}\right], \nn
T&\equiv& \frac12\left[L_{67}+L_{89}\right].
\eeqn
Different notations for the Hamiltonian $H$ and angular momenta
$L's$ are used to emphasize that we are now considering ${\cal N}=8$
theories. See below for more detailed discussion of angular momenta
and how they extends when we includes more hypermultiplets.

$L_{\mu\nu} \;(\mu,\nu=1,2,3,4)$ and $S$ constitute the generators
of the R-symmetry $SO(4)_{1,2,3,4}\times SO(2)_{67-89}$. 
The other generator $T$, on the other hand, does
not correspond to an R-symmetry in ${\cal N}=8$ language. Instead it rotates 
the entire hypermultiplet as a whole  by a $U(1)$ phase. In particular 
$T$ will drop away if we remove all hypermultiplets.
We summarize below how various fields are charged under $(S,T)$,
\beqn
\begin{array}{cccc}
&& S & T \\
X_\mu &\rightarrow & 0 &0 \\
\lambda_1=\lambda &\rightarrow & -1/2 & 0\\
\lambda_2=-i\tilde B\lambda^* &\rightarrow & +1/2 & 0\\
y_1 &\rightarrow & +1/2 &-1/2 \\
y_2 &\rightarrow & -1/2 &-1/2 \\
\chi &\rightarrow & 0& -1/2
\end{array}
\eeqn
For more general theories with other hypermultiplets the same algebra
hold as long as we properly extend the operators to include
these additional matter fields. Two $SO(2)$ global charges of the
general hypermultiplets follows those of adjoint hypermultiplet;
\beqn
\begin{array}{cccc}
&& S & T \\
q_1 &\rightarrow & +1/2 &-1/2 \\
q_2 &\rightarrow & -1/2 &-1/2 \\
\psi &\rightarrow & 0& -1/2
\end{array}
\eeqn
Finally $L_{\mu\nu}$ with $\mu,\nu=1,2,3,4$ which rotates $X^{1,2,3,4}$
acts on all fermions universally via $\Gamma_{\mu\nu}=\gamma_{\mu\nu}\otimes 
1\otimes 1\otimes 1$.

Actually, there is only one independent, 4-component supercharge 
$Q_\alpha\equiv Q^1_\alpha$, since  $Q^2 = -i\tilde B ((Q^1)^\dagger)^T$.
For the sake of completeness, we write down an explicit form of $Q^\dagger$
with one adjoint hypermultiplet. Introducing new notation $\pi_i$ for the
conjugate momenta of $\bar y_i$'s such that 
$[\pi_i,\bar y_j]=-2i\delta_{ij}$, we have
\beqn
Q_\alpha^\dagger&=& \lambda_1^\dagger \left(\gamma_\mu P_\mu 
-\frac{i}{2}[X_\mu,X_\nu]\gamma^{\mu\nu} +i\frac{\mu}{3}X^5
-i\frac{\mu}{6}\sum_{\mu=1}^4X_\mu\gamma^\mu\gamma^5
\right) \nn
&+& i\lambda_1^\dagger 
\left(\frac{[\bar y_2,y_2]-[\bar y_1,y_1]}{2}\right)
-i\lambda_2^\dagger [\bar y_2, y_1]  \nn
&+&\chi_1^\dagger \left( -\pi_1 -i[X_\mu,y_1]\gamma^\mu 
+ i\frac{\mu}{6}y_1\gamma^5\right) \nn
&+&\chi_2^\dagger \left( -\bar \pi_2 -i[X_\mu,\bar y_2]\gamma^\mu 
- i\frac{\mu}{3}\bar y_2\gamma^5\right) .
\eeqn
One must keep in mind that 
\beqn
\lambda_1=\lambda, && \lambda_2=-i\tilde B\lambda^*, \nn
\chi_1=\chi, &&\chi_2=+i\tilde B\chi^*.
\eeqn
Adding more hypermultiplets is done straightforwardly
by copying the hypermultiplet part (last 3 lines) and 
changing the gauge indices. Then, the above superalgebra may 
be rewritten without the symplectic indices as
\beqn
\{{Q}_\alpha,{Q}_\beta\}&=&0, \nn
\{{Q}_\alpha,{Q}_\beta^\dagger\}&=&\delta_{\alpha\beta}2 {H} 
+\mu(\gamma_5)_{\alpha\beta}\, [\,T\,]
+\frac{\mu}{3}(\gamma_5)_{\alpha\beta}\, [\,S\,]
+\frac{\mu}{6}\sum_{\mu,\nu=1}^4
i(\gamma^{\mu\nu}\gamma^5)_{\alpha\beta}\,\left[\,L_{\mu\nu}\,\right] \nn
&+& \sum_{\mu=1}^5 i(\gamma^\mu)_{\alpha\beta}\,
\Tr\left[X^\mu{\cal G}\right],
\eeqn 
and
\beqn
{}[H,Q^\dagger_\beta]&=& \frac{\mu}{12}
Q^\dagger_\alpha(\gamma^5)_{\alpha\beta}.
\eeqn

\section{DLCQ of Fivebranes in PP-Wave Background and Giant Gravitons}

In ordinary M(atrix) theory, $k$ longitudinal fivebranes can be studied
by considering ${\cal N}=8$ SYQM with $k$ fundamental hypermultiplets. As is
well known, this comes about because longitudinal fivebranes are actually
D4 branes whereby D0-D4 strings are introduced. The fivebrane is one of
more difficult objects to understand, largely because their mutual interaction
is governed by tensorial theory instead of usual gauge theories.

In fact, even some of more elementary aspect of fivebrane dynamics are 
still mysterious, such as its number of degrees of freedom. The latter
has been estimated to scale as $k^3$ for large $k$, both from study of 
holography \cite{klebanov} and from  study of conformal \cite{henningson1} 
and axial \cite{hmm} anomalies. No compelling, microscopic explanation of
this counting is available at the moment. One of 
more concrete proposal for study of fivebrane is via DLCQ approach,
which really boils down to study of the above mentioned M(atrix) theory
with fundamental hypermultiplets. Here again, even some of more elementary
questions prove very difficult to answer. For instance, one may ask
quantum vacuum structure of such SYQM, but even this has
not been answered thoroughly. This was answered in some special cases in
Refs.~\cite{H,Stern,Chanju,Arjan,NC}.

One may hope that simplification from the above mass deformation 
will give us some extra handle on fivebranes, in much the same
way that we are hoping to learn more about string theory by studying
its behavior in the pp-wave backgrounds, such as the vacuum structure.
With $k$ fivebranes
lying along $x^{0,6,7,8,9,10}$, in addition to the fivebranes
that gave rise to the pp-wave background, dynamics of $N$ D0 partons
are described by the above deformed ${\cal N}=8$ SYQM with one
adjoint hypermultiplet and $k$ fundamental hypermultiplets.
In this we wish to explore supersymmetric vacua of such a theory.

\subsection{Fuzzy Spheres as Giant Gravitons in the Bulk}

Most intriguing objects found to date in the mass deformed ${\cal N}=16$
M(atrix) theory of BMN are the fuzzy sphere solutions. This consists 
of three nontrivial matrix $X^a$ with $a=5,6,7$ such that
\beq
[X^a,X^b]=-\frac{i\mu}{3}\epsilon_{abc}X^c ,
\eeq
is satisfied \cite{bmn}. 
This can be rescaled to the standard $su(2)$ algebra.
Remarkably, such nontrivial solutions are actually of zero energy and 
preserves all dynamical supersymmetries.

The solutions are classified into irreducible representations of $su(2)$,
so, for $SU(N)$ SYQM, there are as many such solution as the number of
inequivalent partitions of $N$. These classical solutions are strongly
reminiscent of D0 bound states of original M(atrix) theory \cite{Yi}. 
Indeed, the bound states are suppose to exist for any number of D0's, which 
means that, for $SU(N)$ SYQM and $ N=\sum_p N_p$, there are sectors 
with groups of $N_p$ D0 particles bound together. In the current 
mass deformed version, translational degrees of freedom are lost,
and everything is confined near the origin. Nevertheless it is tantalizing 
that hint of these D0 bound states still survives in the form of 
giant gravitons \cite{gg} trapped by the confining potential.

Recall that deformed matrix theory of BMN admits a set of
classical vacua that preserve all dynamical supersymmetries. These
zero energy solutions can  be understood easily from the form of 
potential with $X^i=0$ for $i=1,2,3,4,8,9$.
\beqn
V\rightarrow \Tr \left(-\frac{1}{4}\sum_{a,b=5,6,7}[X^a, X^b]^2 
+ \frac12 \left(\frac{\mu}{3}\right)^2 \sum_{a=5,6,7} (X^a)^2 
-i\frac{\mu}{3} \epsilon_{abc} X^a X^b X^c \right), \label{V}
\eeqn
which can be made into a complete square,
\beqn
-\frac14 \sum \Tr\left( [X^a,X^b] +\frac{i\mu}{3} \epsilon_{abc}X^c\right)^2.
\eeqn
A fuzzy sphere,
\beqn 
X^a=-\frac{\mu}{3}J^a ,
\eeqn
with any $N$ dimensional representation $J^a$ of $su(2)$, is a classical
ground state.

In fact this is sufficient to show that these are invariant under all
dynamical supercharges once we note that superalgebra are
of the form
\beqn
\{Q_\alpha,Q_\beta\} =\delta_{\alpha\beta} 2H +\hbox{angular momentum},
\eeqn
which, for purely bosonic configurations with zero momentum, reduces to 
\beqn
\{Q_\alpha,Q_\beta\} =\delta_{\alpha\beta} 2V .
\eeqn
Thus a purely bosonic solution with $V=0$ and no momentum preserves all 
dynamical supersymmetries.\footnote{More general class of 
solutions that preserves part of the supersymmetries 
may be found \cite{pp}.}

Given $U(N)$ matrix theory, there is precisely one such
distinct state for each partition $N=\sum_p  N_p$, corresponding to the
direct sum of $N_p$ dimensional irreducible representations of $su(2)$.
These classical states and its quantum counterpart is closest thing to a
usual Kaluza-Kline modes of supergraviton, or equivalently BPS bound states
of D-particles. In fact, one may think each of these $N_p\times N_p$ blocks
as the bound state $N_p$ D-particles that are blown-up into a spherical
membrane due to Myers' dielectric effect~\cite{myers}.

\subsection{BPS equations for ${\cal N}=8$}

It thus becomes of some importance to 
search for such supersymmetric classical and quantum vacua in the
deformed matrix quantum mechanics with ${\cal N}=8$ supercharges. In the
simplest case of  pure ${\cal N}=8$ quantum mechanics with no hypermultiplet at
all, adjoint or not, since Myers' cubic term is absent in this case, and
rest of bosonic potential are positive definite, we can immediately see
that no analogous classical solution exists. With one adjoint hypermultiplet, 
we basically come back to BMN case, where the only difference now is that
instead of $X^{a=5,6,7}$, we use the notation $X^5\rightarrow X^5$ and 
$X^6+iX^7 \rightarrow \bar y_2$;
\beqn
&& -\frac14 \Tr\left( [X^a,X^b] +\frac{i\mu}{3} \epsilon_{abc}X^c\right)^2
\label{square0}\\
&& \n\\
\rightarrow 
&& +\frac12 \Tr \left( \frac12[\bar y_2,  y_2] +\frac{\mu}{3} X^5\right)^2 \n\\
&&-\frac12 \Tr\left( ([X^5,\bar y_2] +\frac{\mu}{3} \bar y_2 )( 
[X^5,y_2] -\frac{\mu}{3} y_2)\right) . \label{square1}
\eeqn
It is instructive to keep $y_1$ as well in the potential and write it in 
complete squares,
\beqn
&& +\frac12 \Tr \left( \frac12[\bar y_2, y_2]-\frac12 [\bar y_1,y_1] 
+\frac{\mu}{3} X^5\right)^2 \n\\
&& -\frac12 \Tr\left( ([X^5,\bar y_2] +\frac{\mu}{3}\bar y_2 )( 
[X^5,y_2] -\frac{\mu}{3} y_2)\right)  \n\\
&& -\frac12 \Tr\left( ([X^5,\bar y_1] +\frac{\mu}{6} \bar y_1 )( 
[X^5, y_1] -\frac{\mu}{6}  y_1)\right) \n \\
&&  +\frac18 \Tr \left([\bar y_1, y_2][\bar y_2, y_1]\right) ,
\eeqn
which is more suitable for generalization.\footnote{While this may
suggest a new class of ground states where $y_2=0$ instead of $y_1=0$,
nontrivial $y_1$'s which are zero of the potential does not satisfy
hermiticity conditions. No such additional ground state exists.}

Generalization to cases with additional hypermultiplets is obvious. 
Denoting complex scalars $q^{(f)}_i$ of the $f$-th fundamental 
hypermultiplets, let us write the reduced  potential similarly
\beqn
&&+\frac12 \Tr \left( \frac12[\bar y_2,  y_2] 
+ \frac12 \sum_f \bar q_2^{(f)} q_2^{(f)} 
-\frac12[\bar y_1,  y_1] 
- \frac12 \sum_f \bar q_1^{(f)} q_1^{(f)}
+\frac{\mu}{3} X^5\right)^2   \n\\
&& -\frac12 \Tr\left( ([X^5,\bar y_2] +\frac{\mu}{3} \bar y_2)
([X^5,y_2] -\frac{\mu}{3} y_2 )\right)   \n\\
&& -\frac12 \Tr\left( ([X^5,\bar y_1] +\frac{\mu}{6} \bar y_1)
([X^5,y_1] -\frac{\mu}{6} y_1 )\right)   \n\\
&&+\frac12 \sum_f \Tr\left( (X^5\bar q_2^{(f)} +\frac{\mu}{3} \bar q_2^{(f)})
(q_2^{(f)}X^5 +\frac{\mu}{3} q_2^{(f)} )\right)   \n\\
&&+\frac12 \sum_f \Tr\left( (X^5 \bar q_1^{(f)} +\frac{\mu}{6} \bar q_1^{(f)})
(q_1^{(f)}X^5 +\frac{\mu}{6} q_1^{(f)} )\right)  \n \\
&&+\frac18 \Tr \left(([\bar y_1, y_2]+\sum_f \bar q_1^{(f)} q_2^{(f)})
([\bar y_2, y_1]+\sum_f \bar q_2^{(f)} q_1^{(f)} )\right).
\label{square2}
\eeqn
In the presence of fundamental hypermultiplets, the equation for 
static classical vacua is then
\beqn
-\frac{\mu}{3} X^5 &=& \frac12 [\bar y_2,  y_2]+\frac12 \sum_f 
\bar q_2^{(f)} q_2^{(f)} -\frac12  [\bar y_1, y_1]-\frac12 \sum_f 
\bar q_1^{(f)} q_1^{(f)}  ,\label{D3}\\
{}[X^5, \bar y_1] & =& -\frac{\mu}{6}\bar y_1 ,\label{y1}\\
{}[X^5, \bar y_2] & =& -\frac{\mu}{3} \bar y_2, \label{y2}\\
X^5 \bar q_1^{(f)}  &=& -\frac{\mu}{6}  \bar q_1^{(f)},\label{q1}\\
X^5 \bar q_2^{(f)} &=& -\frac{\mu}{3} \bar q_2^{(f)},\label{q2}\\
0&=&[\bar y_1, y_2]+\sum_f \bar q_1^{(f)}q_2^{(f)}. \label{D12}
\eeqn
These conditions can also be derived from supersymmetry conditions:
these are precisely condition for a purely bosonic solution to preserve
all 8 dynamical supersymmetries. Eq.~(\ref{D3}) and Eq.~(\ref{D12}) follow
from $\delta\lambda=0$ upon using equation of motion for the auxiliary
fields, ${\bf D}$, while middle ones follow from $\delta \chi=0$
and $\delta\psi=0$ respectively.

\subsection{Giant Gravitons on Fivebranes}

Nontrivial solutions with $q=0$ are of course the fuzzy sphere solutions
again, and corresponds to states in the ``bulk''. 
In the presence of fivebranes, new supersymmetric solutions arise with 
$q\neq 0$. A particularly simple set of solutions can be found in Abelian
case, corresponding to a single D0 probing $k$ fivebranes at origin
in the pp-wave limit. $X^5$ could be either $-\mu/3$ for $q_1=0$ or 
$-\mu/6$ for $q_2=0$. However, nonnegativity of $\bar q_i q_i$ disallow the
latter solution, so we must take $q_1=0$ and $X^5=-\mu/3$. Commutators 
vanish identically, implying that $y_i=0$ also, and the equations simplify to
\beqn
\frac12 \sum_f |q_2^{(f)}|^2 =-\frac{\mu}{3} X^5 =
\left(\frac{\mu}{3} \right)^2 .
\eeqn
Thus, a continuous set of solutions exist and are parametrized as
\beq
q^{(f)}_2=\sqrt2 \frac{\mu}{3} z^f ,
\eeq
where $z$ is any complex $k$-vector of unit length. Furthermore,
we must identify vacua up to $U(1)$ gauge rotation, and so
the vacuum moduli space is $CP^{k-1}$. Quantization of this
will lead to a unique supersymmetric ground state, since the
low energy dynamics is purely bosonic, and should be related to 
the bound states found in Ref.~\cite{Chanju}.

For nonabelian theories again with $k$ hypermultiplet, there is
a direct generalization of the above family of solutions. Take
an $N\times N$ block of adjoint scalars and write 
\beq
-(X^5)_{AB}=\frac{\mu}{3}(n+1-A)\delta_{AB},
\eeq
where $(N-1)/2< n\le N$ so that $(X^5)_{nn}=-\mu/3$. 
With $q_1=0=y_1$, the following form of  $\bar q_2$ and
$\bar y_2$ solve the BPS equations
\beq
(\bar q_2^{(f)})_A=\sqrt2\frac{\mu}{3}w^f_2 \delta_{A,n}, 
\qquad (\bar y_2)_{AB}=\sqrt2\frac{\mu}{3}\alpha_A \delta_{A,B-1}
\eeq
with
\beqn
|\alpha_1|^2 &=& n  , \n\\
|\alpha_2|^2 -|\alpha_1|^2 &=& n-1  , \n\\
&\cdot &   \n\\
&\cdot &   \n\\
|\alpha_{n-1}|^2 -|\alpha_{n-2}|^2 &=& 2  , \n\\
\sum|w^f_2|^2+|\alpha_n|^2 -|\alpha_{n-1}|^2 &=& 1 , \n\\
|\alpha_{n+1}|^2 -|\alpha_{n+2}|^2 &=& 0  , \n\\
&\cdot &   \n\\
&\cdot &   \n\\
|\alpha_{N-1}|^2 -|\alpha_{N-2}|^2 &=& n+2-N , \n \\
-|\alpha_{N-1}|^2 &=& n+1-N .
\eeqn
Note that this family of solution is again parametrized by $CP^{k-1}$.

Another set of solutions are
\beq
-(X^5)_{AB}=\frac{\mu}{3}(m+1/2-A)\delta_{AB},
\eeq
where $1\le m < N/2 $ so that $(X^5)_{mm}=-\mu/6$. 
With $q_2=0=y_1$, the following form of  $\bar q_1$ and
$\bar y_2$ solves the BPS equation,
\beq
(\bar q_1^{(f)})_A=\sqrt2\frac{\mu}{3}w_1^f \delta_{A,m}, 
\qquad (\bar y_2)_{AB}=\sqrt2\frac{\mu}{3}\alpha_A \delta_{A,B-1},
\eeq
with
\beqn
|\alpha_1|^2 &=& m-1/2  , \n\\
|\alpha_2|^2 -|\alpha_1|^2 &=& m-3/2  , \n\\
&\cdot &  \n\\
&\cdot &   \n\\
|\alpha_{m-1}|^2 -|\alpha_{m-2}|^2 &=& 3/2  , \n\\
-\sum |w^f_1|^2+|\alpha_m|^2 -|\alpha_{m-1}|^2 &=& 1/2 , \n\\
|\alpha_{m+1}|^2 -|\alpha_{m+2}|^2 &=& -1/2  , \n\\
&\cdot &   \n\\
&\cdot &   \n\\
|\alpha_{N-1}|^2 -|\alpha_{N-2}|^2 &=& m+3/2-N , \n \\
-|\alpha_{N-1}|^2 &=& m+1/2-N .
\eeqn
Each integer $m$ ($1\le m < N/2 $) or $n$ ($(N-1)/2< n\le N$)   
gives distinct sets of solutions, which are each
parametrized by $CP^{k-1}$.

\section{Conclusion}

In this note, we derived massive deformation of ${\cal N}=8$ supersymmetric
Yang-Mills quantum mechanics, analogous to BMN's deformation of M(atrix) 
theory. A particular case with one adjoint hypermultiplet and $k$
fundamental hypermultiplet is interpreted as D0 dynamics in the
presence of $k$ longitudinal fivebranes, in the M-theory pp-wave 
background. A very rich structure of vacua are shown to exist when
fundamental hypermultiplets are present, typically parametrized by
$CP^{k-1}$ vacuum moduli spaces.

Clearly these states are deformation of fuzzy spheres, or giant gravitons,
of ${\cal N}=16$ deformed M(atrix) theory of BMN, where deformation entails
turning on fundamental hypermultiplets. It is tempting to view these
states, since they protrude into the Higgs phase, as giant gravitons 
``trapped'' by fivebranes. However, it remains largely mysterious 
what role they may play in dynamics of fivebranes. 

Another aspect that needs be addressed is the matter of quantum ground 
states, as opposed to classical ground states. Proliferation of classical 
ground states will be in part lifted once quantum consideration is used;
for each $CP^{k-1}$ vacuum moduli space, there is a unique quantum
ground state, almost certainly. Still there are probably more than one 
inequivalent quantum ground states, given $k\ge 1$ and $N>1$, since vacuum 
manifold consists of many disconnected $CP^{k-1}$. The matter of quantum
spectra of these ${\cal N}=8$ theories will be investigated elsewhere.

Before closing, it is worthwhile to note that the kind of reduction 
process we have taken can be repeated for ${\cal N}=8$ theory. 
For instance, start with the deformed ${\cal N}=8$ pure SYQM of section 2.3,
truncate away $X^4$, $X^5$ while restricting $\lambda$ to be
chiral under $\gamma^5$. At the end of the day, one obtains a pure
${\cal N}=4$ SYQM which is mass deformed by $\mu$. Again for this theory,
cancellation of vacuum energy is complete,
\beq
3\times \frac{\mu}{6} -2 \times \frac{\mu}{4} =0,
\eeq
since, in each vector multiplet, there are 3 bosons of mass $\mu/6$
and 2 complex fermions of mass $\mu/4$.
At the moment it is unclear to us what is a pp-wave interpretation, 
and indeed whether  there is one. Regardless
of stringy interpretation, however, such large class of 
mass-deformed SYQM may provide 
further insights on issues on vacuum structure of general SYQM.

\subsection*{Acknowledgement}
This work is supported in part by  KOSEF 1998 interdisciplinary research 
grant 98-07-02-07-01-5 (K.L.). N.K. is grateful to J.-H. Park for useful
comments and discussions, and acknowledge the hospitality of KIAS 
where part of the work was done.

\end{document}